\documentstyle[amssymb,prb,aps,twocolumn]{revtex}

\begin{document}
\draft
\title{Spin Dynamics of Rare Earth Ions in Phosphate Laser Glasses}
\author{I.P. Goudemond, J.M. Keartland, M.J.R. Hoch,}
\address{Department of Physics, University of the Witwatersrand,\\
P O WITS 2050, South Africa}
\author{and G.A. Saunders}
\address{School of Physics, University of Bath, \\
Claverton Down, Bath BA2 7AY, UK}
\date{\today }
\maketitle

\begin{abstract}
The spin dynamics of rare earth ions (Er, Nd, Sm and Gd) in heavily doped
phosphate glasses have been investigated using $^{31}$P NMR. Correlation
times in the range 10$^{-2}$ to 10$^{-12}$ s have been obtained for
temperatures between 4 K and 100 K. No evidence of spin-spin coupling
between the ions has been found and spin relaxation occurs via conventional
phonon processes, including the Orbach process.

A model involving inhomogeneous broadening of the NMR resonance lines, with
distinct sample regions corresponding to the presence or absence of nuclear
spin diffusion, has been used in extracting the electron spin correlation
times from the NMR measurements.
\end{abstract}


\section{Introduction}

Novel magnetic and magneto-optic phenomena, of interest for optoelectronic
applications, have been found in the rare earth (R) metaphosphate glasses
(REMG) with compositions in the vicinity (R$_{2}$0$_{3}$)$_{0.25}$(P$_{2}$0$%
_{5}$)$_{0.75}$. For example, these paramagnetic materials exhibit the
largest known magnetic contributions to the low temperature specific heats
in oxide glasses $^{\text{1}}.$ X-ray diffraction and EXAFS studies$^{2-4}$
have shown that the structure of REMG comprises a 3-D network of
corner-linked PO$_{4}$ tetrahedra, with the rare-earth ions, which in
several instances are known to be trivalent R$^{3+},$ occupying sites within
the PO$_{4}$ skeleton.

In recent work $^{5},$ we have shown that it is possible to study the spin
dynamics of rare earth ions in metaphosphate glasses using $^{31}$P NMR as a
probe. The present investigation has extended the NMR measurements to a
number of REMG systems containing Er, Nd, Sm and Gd ions. For some of these
systems, La or Y ions have been used as a buffer in order to lower the
concentration of magnetic ions while preserving the metaphosphate glass
structure. Information on the dynamics of the rare earth ions has been
obtained using the model developed previously $^{5}$.

\section{Theory}

The local fields set up by paramagnetic ions lead to nuclear resonance
frequency shifts and the establishment of an exclusion barrier of radius $%
\rho _{c}$, inside which spins are excluded from the observed NMR signal,
together with a diffusion barrier of radius $b_{0},$ inside which nuclear
spin diffusion does not operate$^{5}$. In highly paramagnetic systems of the
type studied here, nuclear spin systems are thus composed of three distinct
spatial regions corresponding to spin diffusive, non-diffusive and exclusion
zones. The theory of nuclear spin lattice relaxation (NSLR) due to
localized, isolated paramagnetic centres in magnetically dilute crystals$%
^{6-11}$ has been extended to account for magnetically concentrated systems$%
^{5}$. In such systems, nuclear spin diffusion may be inoperative in
appreciable fractions of sample volume, and nuclei may relax to multiple
paramagnetic sites. The average relaxation rate for nuclei in diffusive
regions is given by 
\begin{equation}
{\displaystyle {1 \over T_{1}}}%
=\frac{4\pi }{3}n_{s}C^{\frac{1}{4}}D^{\frac{3}{4}}  \label{DL}
\end{equation}
in the diffusion limited [DL] case, and 
\begin{equation}
{\displaystyle {1 \over T_{1}}}%
=\frac{4\pi }{3}\frac{n_{s}\gamma _{I}^{\frac{3}{2}}}{(\gamma _{I}\gamma
_{S}\hbar )^{\frac{3}{2}}S^{\frac{3}{2}}}C\left( \frac{\Delta \omega }{%
\omega _{n}}\right) ^{\frac{3}{4}}\left( \frac{k_{B}T}{3a_{0}}\right) ^{%
\frac{3}{4}}  \label{RD}
\end{equation}
in the rapid diffusion [RD] case, with 
\begin{equation}
C=\frac{2}{5}(\gamma _{I}\gamma _{S}\hbar )^{2}S(S+1)\frac{\tau }{1+\omega
_{n}^{2}\tau _{e}^{2}}.  \label{C}
\end{equation}
$D$ is the nuclear spin diffusion coefficient, $n_{s}$ is the magnetic ion
concentration and $\omega _{n}$ is the nuclear Larmor frequency$^{5}$.

The net electronic relaxation rate for ions with a crystal field splitting
includes direct, Raman and Orbach relaxation rates$^{12-14}$:

\begin{equation}
\frac{1}{\tau _{^{e}}}=\frac{1}{C_{D}\Delta ^{2}}T+\frac{1}{C_{R}\Delta ^{2}}%
T^{7}+\frac{\Delta ^{3}}{C_{O}}\frac{1}{\exp \left( 
{\displaystyle {\Delta  \over k_{B}T}}%
\right) -1}.  \label{tau_e}
\end{equation}

{\it C }$_D${\it , C}$_R${\it \ }and{\it \ C}$_O$ are relaxation constants
for the direct , Raman and Orbach processes, respectively. $\Delta $ is the
crystal field splitting.

In Gd$^{3+}$ doped glasses, there is no crystal field splitting, and the
ground state is a multiplet, so the net relaxation rate may be expressed as

\begin{equation}
\frac{1}{\tau _{^{e}}}=\frac{1}{C_{D}}T+\frac{1}{C_{R}}T^{5}.  \label{tau_gd}
\end{equation}

Equations \ref{tau_e} and \ref{tau_gd} assume a Debye phonon density of
states. This is consistent with recent work$^{15},$ including the soft
potential model approach, which suggests that vibrational modes in these
systems are phonon-like. Fractons$^{17,18}$ and 'excess' modes$^{15}$
involving low-energy excitations do not appear to play a role in the
relaxation processes in these systems.

\section{Experimental Details}

The samples were prepared from the melt at Bath University with original
stoichiometry$^{24}$ (R$_{2}$O$_{3}$)$_{x}$(X$_{2}$O$_{3}$)$_{.25-x}$(P$_{2}$%
O$_{5}$)$_{.75}$, and $x$ in the range $0.01$ to $0.25$ (nominally), for R =
Er$^{3+}$, Nd$^{3+}$, Sm$^{3+}$ and Gd$^{3+}$, and X = Y$^{3+}$ or La$^{3+}$.

$^{31}$P$\ $NMR measurements at 8.5MHz (0.49 T) and 19.25MHz (1.1 T) were
carried out using a pulsed NMR spectrometer and a Varian water-cooled
electromagnet. Preliminary EPR measurements were carried out on Er and Gd
REMG using a commercial X-band Bruker ESP380E pulsed EPR spectrometer and an
Oxford continuous-flow helium cryostat to achieve low temperatures.

Nuclear resonance lineshapes have been characterised in terms of the moments
of the resonance line determined from the Fourier Transform frequency-domain
NMR spin-echo waveforms for samples with large inhomogeneous broadening.

The average nuclear relaxation rates for diffusive regions were extracted
from the magnetization recovery data as a function of time $\tau $ using$%
^{5} $ 
\begin{equation}
\frac{M(\tau )}{M_{0}}=1-2\nu \left[ f\cdot \exp \left( \frac{-\tau }{T_{1}}%
\right) +(1-f)\cdot \exp \left( \frac{-\tau }{\lambda T_{1}}\right) ^{\frac{1%
}{2}}\right] ,  \label{relax}
\end{equation}

where $\nu $ is a scaling factor, introduced because the degree of nuclear
saturation is not known precisely, and $f$ is the diffusive fraction of
nuclei. The first and second terms describe NSLR in diffusive and
non-diffusive regions, respectively. The relaxation rate for non-diffusive
regions has been expressed in the form $T_{1}^{\prime }=\lambda T_{1}$ for
convenience, where $\lambda $ is the ratio of the average relaxation rates
for nuclei in diffusive and non-diffusive regions.

\section{Results and Discussion}

We were unable to observe EPR free induction decay or spin-echo signals in
the samples investigated, probably due to extremely short spin-spin
relaxation times in these systems. Direct measurements of the electron
spin-lattice relaxation time using pulsed EPR techniques have, therefore,
not been possible.

Nuclear resonance lineshapes have been characterised in terms of the moments
of the resonance line, using $\Delta \omega =<M_{2}>^{\frac{1}{2}}$, and the
moment ratio, $R=<M_{4}>/<M_{2}>^{2}$. The NMR resonance line is
inhomogeneously broadened due to coupling of the nuclear and paramagnetic
ion moments. Fig. 1a shows plots of the inhomogeneouosly broadened $^{31}$P
nuclear resonance linewidth and moment ratio $R$ versus inverse temperature
for a 1\% Er REMG at B = 1.1 T. The linewidth increases with decreasing
temperature, eventually reaching a plateau at approximately 20 K. The ratio 
{\it R} is close to 3 at low temperatures, suggesting a characteristically
Gaussian lineshape. The natural dipolar linewidth is estimated to be
approximately 3.8 kHz, so there is appreciable inhomogeneous broadening of
the linewidth even at the highest temperatures at which measurements were
made. Following the onset of motional narrowing, {\it R} increases to around
5, implying a change in lineshape. This behaviour is typical of all the
glasses studied. Fig. 1b shows plots of linewidth versus inverse temperature
for various Er REMG samples . For each sample, the linewidth increases while
the signal amplitude decreases with decreasing temperature, eventually
becoming unobservably weak over a small temperature range. The
signal-to-noise ratio increases with increasing resonance frequency so that,
for a given sample, the temperature range over which the line can be
observed is somewhat larger for a larger applied field.

It appears that the maximum observed linewidth depends on the magnetic ion
concentration, the particular ion present and the magnetic field. In all
cases, the NMR signal became immeasurably broad over a very small
temperature range at sufficiently low temperatures. The relatively low
maximum observed linewidth ($\Delta \omega $ 
\mbox{$<$}%
\ 25 kHz ) and the difference in maxima for different experiments strongly
suggest that the abrupt loss of signal at low temperatures is not associated
with limited spectrometer bandwidth. This loss of signal appears to be
linked to the overlapping of diffusion barriers when two or more ions
contribute to the line-broadening process for a majority of nuclear spins in
all regions of the sample.

The $^{31}$P spin-spin relaxation time, $T_{2},$ determined from the
spin-echo envelope, is 260 $\mu $s. For the Er$^{3+}$, Nd$^{3+}$ and Sm$%
^{3+} $ ion doped glasses, the electron correlation time approaches the
nuclear $T_{2}$ value at temperatures in the vicinity of 5 K. For Gd$^{3+}$,
the temperature at which $\tau _{e}$ becomes comparable to $T_{2}$ is
approximately 50 K. When $\tau _{e}$ $\lesssim $ $T_{2}$, we expect the
diffusion and exclusion barrier radii to reach their maximum values.

Fig. 2a shows plots of $^{31}$P relaxation rates $T_{1}^{-1}$ versus $T^{-1}$
for the 1\% Nd, 6\% Nd and metaphosphate Nd REMG for $B$ = 4.9 kG. The peak
in the relaxation rate for the 1\% Nd data corresponds to $\omega _{n}\tau
_{e}$ = 1. The temperature range over which measurements could be made on
the 6\% Nd and metaphosphate Nd REMG before the line broadened dramatically
rendering the signal unobservable, was insufficient to observe the peak. It
is easily seen that the nuclear relaxation rate is proportional to the
magnetic ion concentration, but it appears that the metaphosphate Nd REMG
contains a magnetic ion concentration lower than the nominal 25\%. The data
have been analysed using Eq.\ref{DL} [DL] and Eq.\ref{RD} [RD]. The curves
represent the best fits to the data of the [RD] and [DL] expressions. The Nd$%
^{3+}$ data are well-described by the [RD] expression above 20 K, and follow
the [DL] expression at lower temperatures. In Nd$^{3+},$ the contributions
to electronic relaxation from the Raman and direct processes are negligible,
and the data were analysed taking only the Orbach process into account for
the temperature range covered in these experiments. All three Nd data sets
were fitted using the same Orbach relaxation parameters. Both the lineshape
data and the relaxation data provide no evidence that spin-spin interactions
between the rare earth ions are important over the temperature range covered
in the present experiment.

Fig. 2b shows plots of $^{31}$P relaxation rates $T_1^{-1}$ versus $T^{-1}$
for the 1\% Nd, 1\% Er, and 20\% Sm REMG. The nuclear relaxation rates for
the Nd and Er REMG are similar at high temperatures. The peak in the Nd data
occurs at a lower temperature than in the Er and Sm data. This suggests
faster electronic relaxation in the Nd$^{3+}$ ions, which results from a
smaller crystal field splitting than in the other REMG. Nuclear relaxation
in the 20\% Sm REMG is much faster than in the Er and Nd REMG because of the
much higher dopant concentration.

For the 1\% Gd REMG, electronic relaxation is much less efficient than in
the other REMG due to the absence of the Orbach process. The condition $%
\omega _{n}\tau _{e}$ = 1 is not satisfied within the temperature range, and
all the data are on the low temperature side of the relaxation peak (Fig.
2c). In Gd$^{3+},$ the whole data set is well described by the [DL]
expression. The direct relaxation process dominates relaxation below 20 K,
while the Raman process is most important at higher temperatures. Electronic
relaxation parameters for Nd$^{3+}$, Er$^{3+}$, and Sm$^{3+}$ ions appear in
Table 1. For Gd$^{3+}$ ions, the Orbach process is not operative and direct
and Raman processes determine the ionic relaxation time. We obtain C$_{\text{%
D}}$ = 8.1(2)$\times $ 10$^{-2}$(s.K) and C$_R$ = 2.9(2) $\times 10^5$ (s.K$%
^5$) for the Gd$^{3+}$ REMG.

The parameters listed in Table 1, together with C$_R$ and C$_{\text{D}}$ for
Gd$^{3+},$ have been used to generate plots of the electronic relaxation
rate $\tau _e^{-1}$ vs. $T^{-1}$ for the magnetic ions studied (Fig. 3). It
can be seen that the relaxation rates for ions with a crystal field
splitting are much higher than for the Gd$^{3+}$ ions. As noted above, it is
clear from Fig. 2b that the relaxation behaviour of the 1\% Nd and 1\% Er
glasses is very similar at higher temperatures. Fig. 3 shows that a
crossover in electron correlation times for these two systems occurs at
around 15 K.

The quality of the fits of the relaxation expressions to the experimental
data and the consistency of the fitting parameters between data sets provide
strong support for the model as well as for the use of electronic relaxation
expressions assuming a phonon density of states.

\section{Conclusion}

Measurements of the $^{31}$P spin-lattice relaxation time in metaphosphate
glasses doped with rare earth ions have provided information on the spin
dynamics of these ions over a wide range of temperature. For Er, Nd and Sm
ions, the observed behaviour with temperature is similar, with the spin
correlation time being determined by a two-phonon Orbach process. The local
environment of the ions is a distorted octahedron of oxygen atoms which give
rise to crystal field splittings. Estimates of the splitting are provided by
the present work for the ions mentioned above. In the case of Gd ions, where
no crystal field splitting occurs, the paramagnetic ion correlation time is
determined by direct and Raman processes. The approach we have established
for determining details of paramagnetic ion behaviour in these systems
should be widely applicable to similar systems.

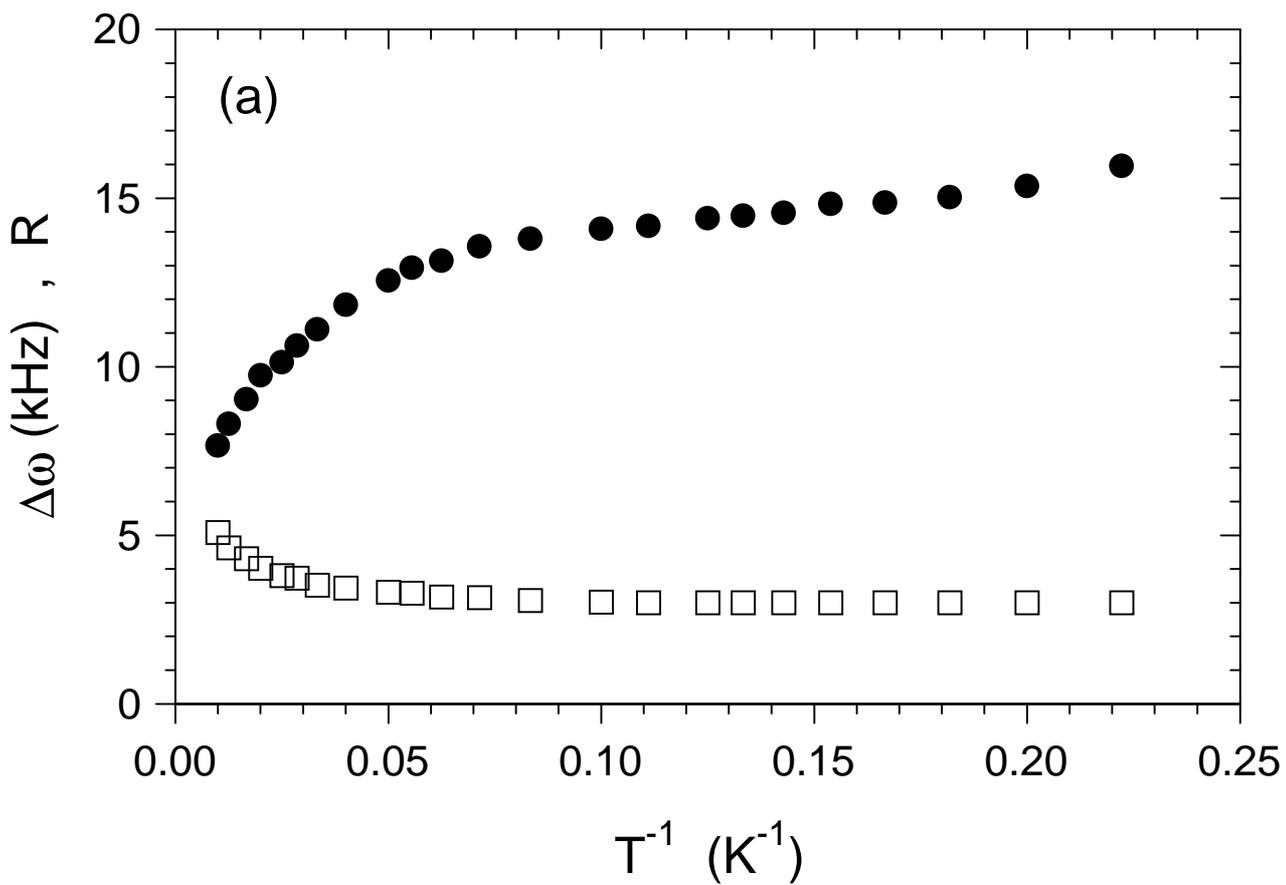
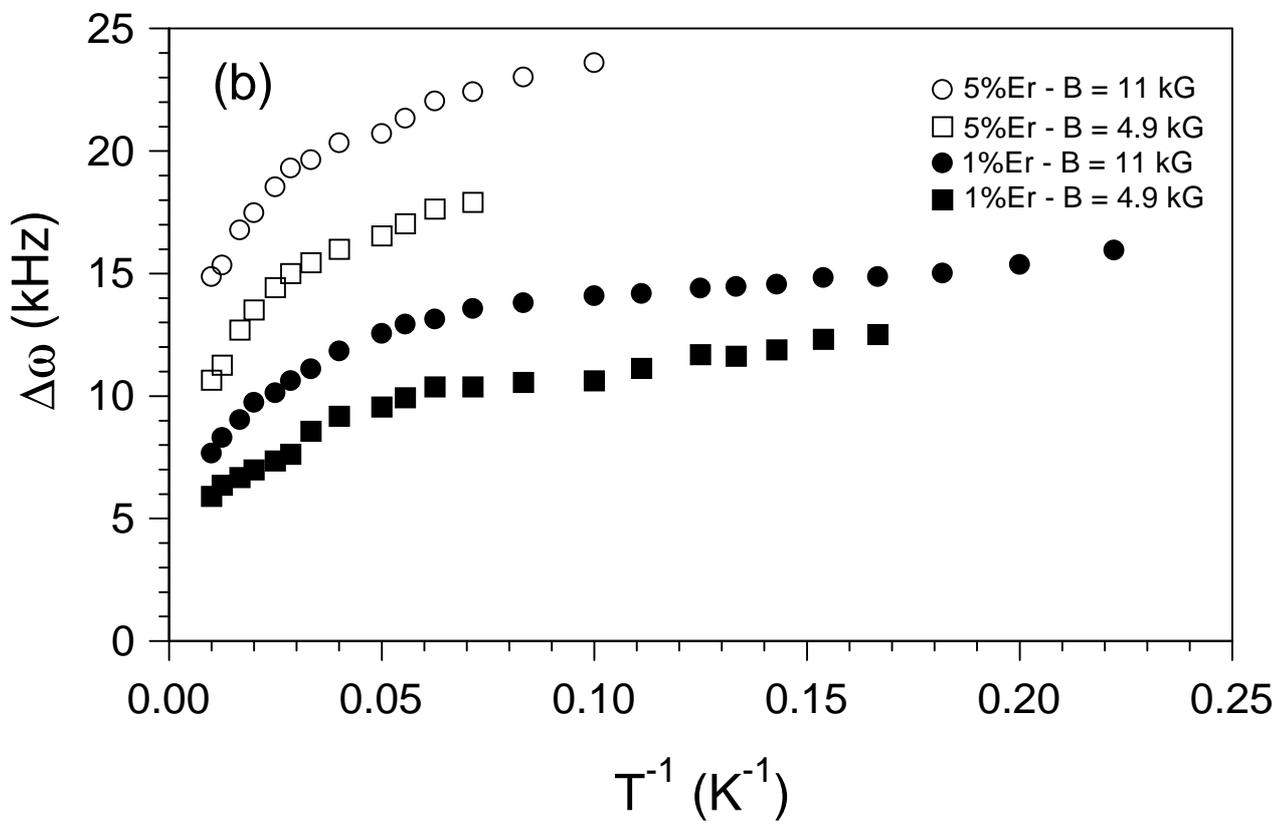
\begin{figure}[tbp]
\caption{(a)Inhomogeneously broadened nuclear resonance linewidth (dots) and
moment ratios (squares) for a 1\% Er REMG with $B$ = 1.1 T.
(b)Inhomogeneously broadened nuclear resonance linewidths for 1\% Er and 5\%
Er REMG with $B$ = 0.49 T and $B$ = 1.1 T.}
\label{Fig.1}
\end{figure}

\begin{figure}[tbp]
\caption{(a)Plots of T$_{1}^{-1}$ versus T$^{-1}$ for 1\% Nd, 6\% Nd and
Metaphosphate Nd REMG. (b)Plots of T$_{1}^{-1}$ versus T$^{-1}$ for 1\% Nd,
1\% Er and 20\% Sm. (c) A plot of T$_{1}^{-1}$ versus T$^{-1}$ for 1\% Gd
REMG. The curves are the best fits to the experimental data of the [RD] and
[DL] expressions.}
\label{Fig.2}
\end{figure}

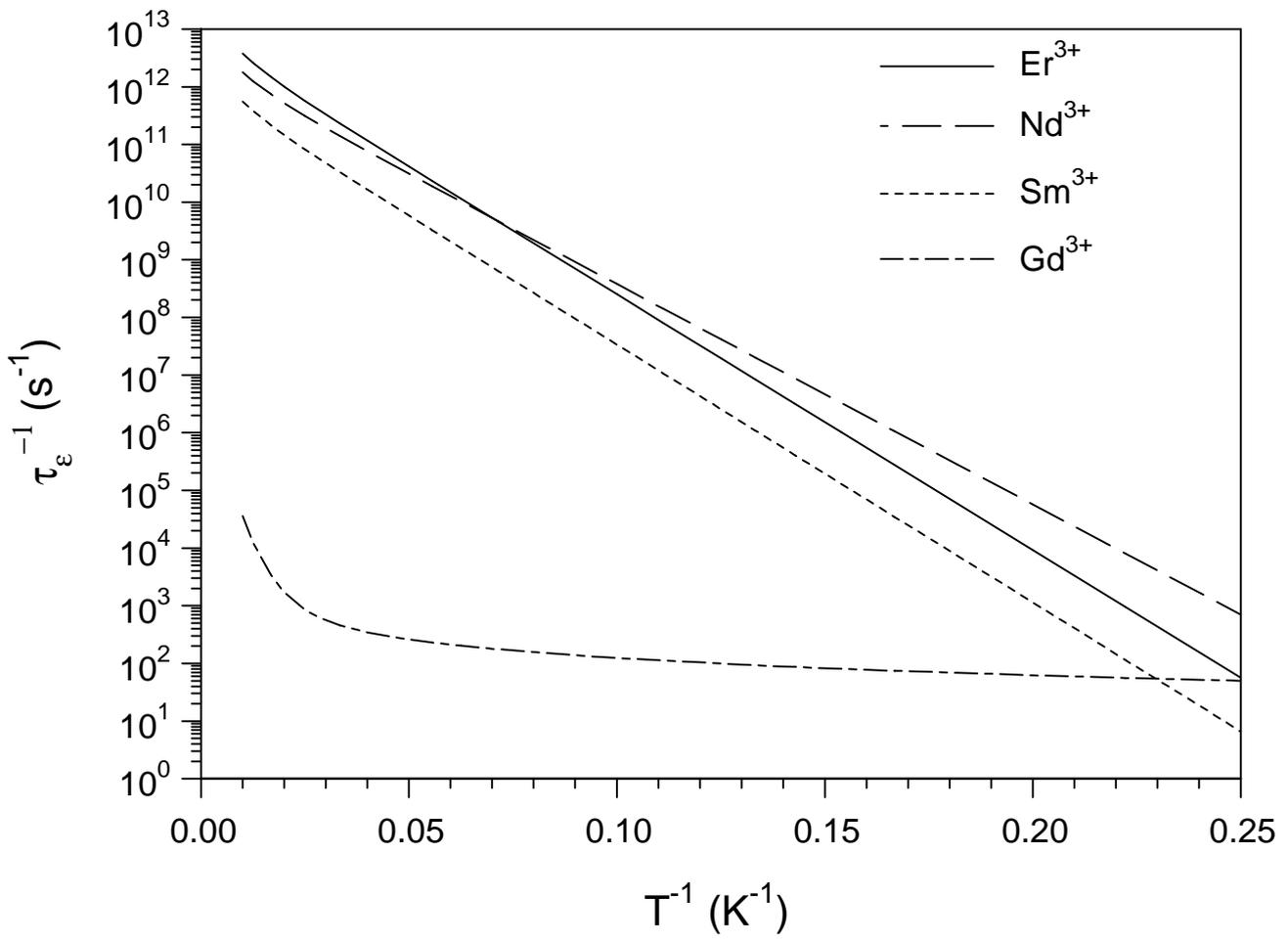
\begin{figure}[tbp]
\caption{Plots of the electronic relaxation rate extracted from nuclear
relaxation data vs. inverse temperature for the magnetic rare earth ions
studied.}
\label{Fig.3}
\end{figure}

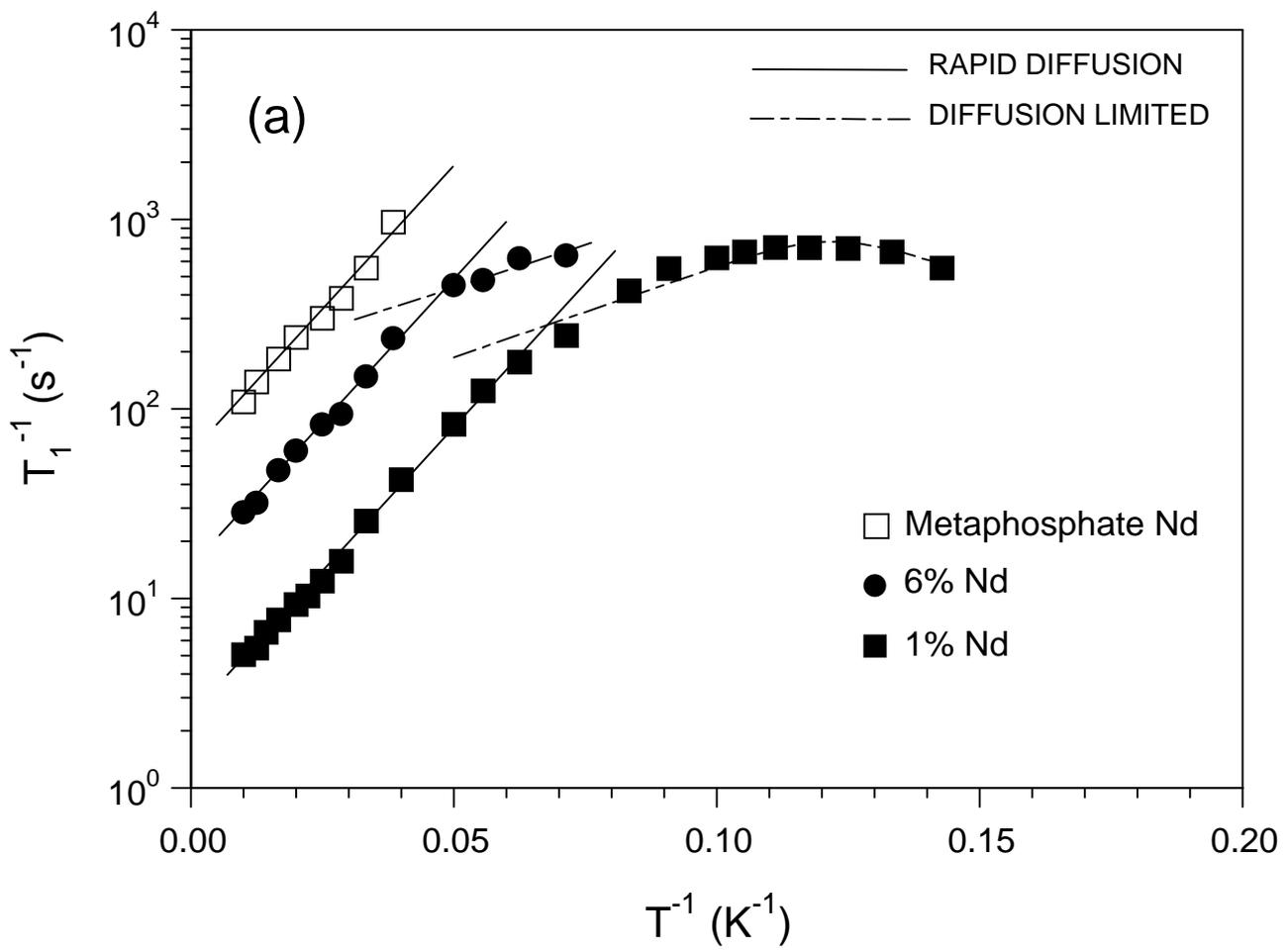
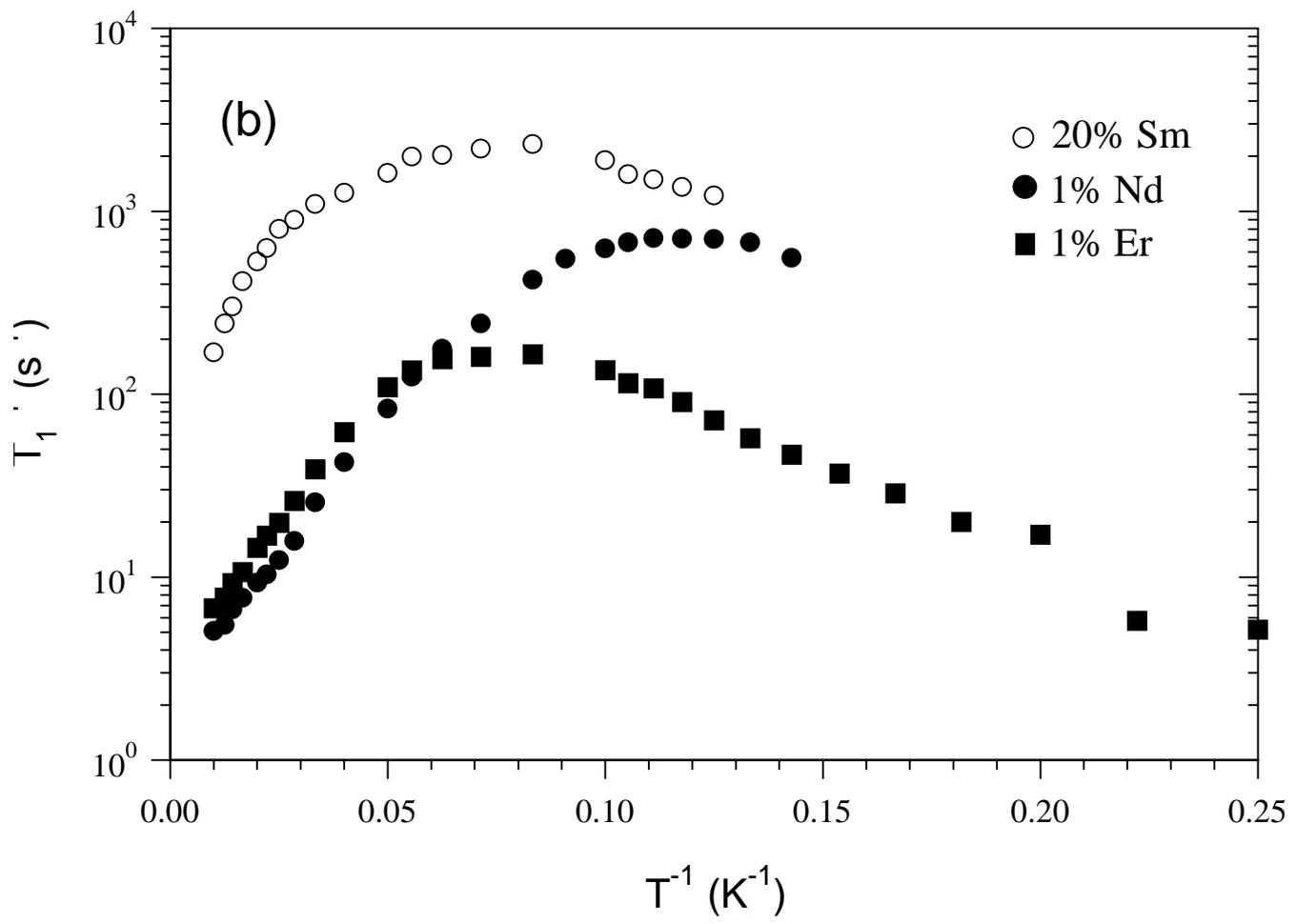
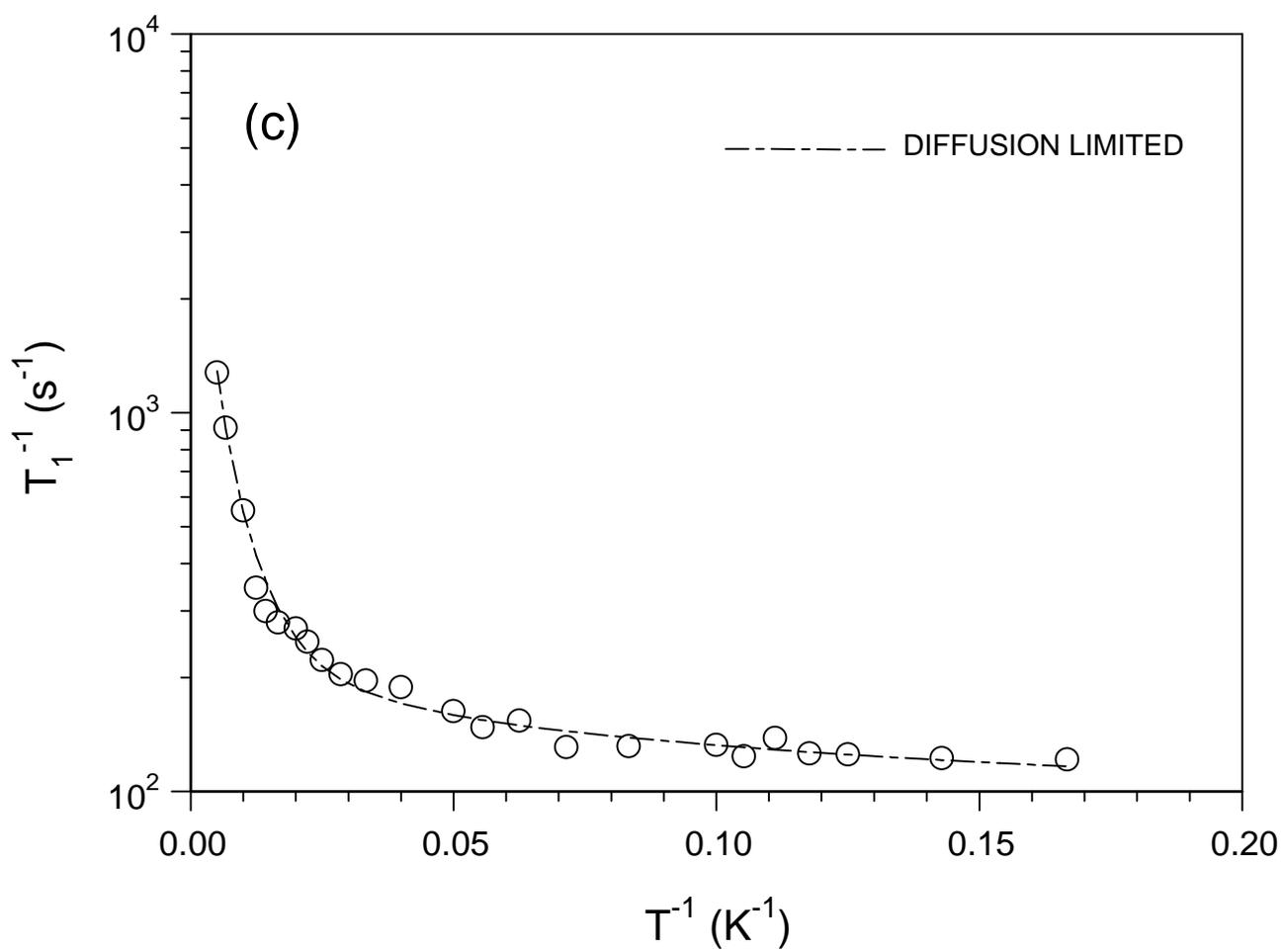
\end{document}